\documentclass[aps,pre,showpacs,twocolumn,groupedaddress]{revtex4-1}
\usepackage{amsmath,amssymb,graphicx,tikz,subfigure}
\usepackage[english]{babel}
\usepackage[utf8]{inputenc}
\usepackage{stmaryrd}
\usepackage{epstopdf}

\newcommand{\bq}{\bar{q}}

\newcommand{\pa}{\partial}

\newcommand{\lan}{\langle}
\newcommand{\ran}{\rangle}
\newcommand{\sign}[0]{\text{sign}}

\newcommand{\ph}{\varphi}
\newcommand{\bp}{\bar{p}}
\newcommand{\ds}{\displaystyle}
\newcommand{\spa}[1]{\qquad\mathrm{#1}\qquad}

\begin{document}

\title{Stiffness of random walks with reflecting boundary conditions}

\author{Sascha Kaldasch}
\author{Andreas Engel}

\affiliation{Universit\"at Oldenburg, Institut f\"ur Physik, 26111 Oldenburg, Germany}

\begin{abstract} We study the distribution of occupation times for a one-dimensional random walk restricted to a finite interval by reflecting boundary conditions. At short times the classical bimodal distribution due to Lévy is reproduced with walkers staying mostly either left or right to the initial point. With increasing time, however, the boundaries suppress large excursions from the starting point, and the distribution becomes unimodal converging to a $\delta$-distribution in the long time limit. An approximate   spectral analysis of the underlying Fokker-Planck equation yields results in excellent agreement with numerical simulations. 
\end{abstract}

\maketitle

\section{Introduction}
Random walks are central to the theory of stochastic processes, both because they can be analyzed analytically in great detail and because they are ubiquitous in nature. Despite their simplicity they have some intriguing properties and exhibit several counterintuitive features. A prominent example is the surprising ``stiffness'' of an unbiased one-dimensional random walk: in a fixed time interval the walker will most of the time stay either left or right to its starting point. Trajectories that stay half of the time on either side of the initial point have the smallest probability in apparent contrast to the fact that jumps to the left and to the right are equally likely. 

\begin{figure}[h]
\centering
\includegraphics[width=.55\linewidth]{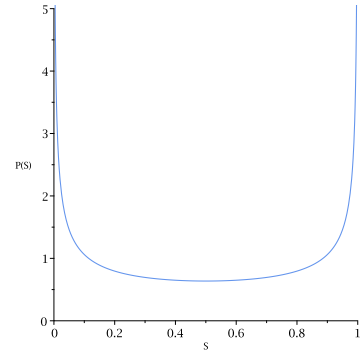}
\caption{Distribution of the fraction of time $S$ an unrestricted symmetric one-dimensional random walker spends to the right of the starting point.}
\label{fig:levy}
\end{figure}

More precisely, the probability density $P_T(S)$ for the fraction $S$ of overall time $T$ the walker spent to the right (or left) of the starting point is given by 
\begin{equation}
 P_T(S)=\frac{1}{\pi\sqrt{S(1-S)}},
\end{equation} 
cf. Fig.~\ref{fig:levy}. This remarkable result was established long ago by Paul Lévy \cite{Levy} and gave rise to a plethora of discussions and generalizations in the mathematics and physics literature, see, e.g.,~\cite{FellerI,FellerII,Yor,Redner,Majumdar} and references therein. Examples include the generalization to cases with deterministic \cite{Akahori,Dassios} and random drift fields \cite{MaCo} as well as to anomalous diffusion \cite{Barkai}, the investigation of different large deviation properties of $P_T(S)$ \cite{MaBr,Barkai2,BuTo}, and the possibility of dynamical phase transitions \cite{NyTo1,NyTo2}.

From an intuitive point of view one may suspect that the large values of $P_T(S)$ near $S=0,1$ arise because the walker may travel arbitrarily long distances away from the starting point. This is corroborated by other well-known properties of one-dimensional random walks like the fact that although return to the origin is certain, the mean time for it diverges, by the distribution of the number of returns to the starting point \cite{FellerI}, by the statistics of successive returns \cite{McFa,Redner,Barkai}, as well as by the asymmetry of the random walk \cite{Weiss}. The shape of the distribution $P_T(S)$ should therefore change qualitatively if reflecting boundary conditions restrict the random walk to a finite interval. 

To test this conjecture we analyze in the present paper the distribution $P_T(S)$ for an unbiased one-dimensional random walk with reflecting boundary conditions symmetric to the starting point. Although various properties of random walks on finite intervals have been investigated in the past \cite{FellerIII,Grebenkov,BuTo,Barkai} the details of the shape transformation in $P_T(S)$ have not been elucidated so far. The problem may be mapped onto a Sturm-Liouville eigenvalue problem that cannot completely be solved analytically. Nevertheless, we provide a highly accurate approximate solution that is in perfect agreement with results from numerical simulations. We find that for small $T$ the distribution is still of the form shown in Fig.~\ref{fig:levy} but changes to an unimodal distribution with maximum at $S=1/2$ with increasing duration $T$ of the walk. For $T\to\infty$ it approaches a $\delta$-distribution around $S=1/2$ in accordance with equilibrium statistical mechanics \cite{Barkai}. 

The paper is organized as follows. In section~\ref{sec:be} we introduce the basic notation and establish the central Fokker-Planck equation for the joint probability distribution of the walker position $x$ and the time fraction $S$. In section~\ref{sec:FPE} we show how to map the solution of this equation to a Sturm-Liouville eigenvalue problem, that we analyze in section~\ref{sec:SL}. Section~\ref{sec:num} discusses the numerical determination of the eigenvalues. In section~\ref{sec:asy} we analytically extract the asymptotics of $P_T(S)$ for large $T$. In section~\ref{sec:res} we present results for $P_T(S)$ for various values of $T$ and compare them with numerical simulations. Finally, section~\ref{sec:conc} contains some conclusions. 

%%%%%%%%%%%%%%%%%%%%%%%%%%%%%%%%%%%%%%%%%%%%%%%%%%%%%%%%%%%%%%%%%%%%%%%%%%%%%%%%%%%%%%%%%%%%%%%%%

\section{Basic Equations}\label{sec:be}
We consider the time interval $0\leq t\leq T$ of a one-dimensional random walk $x(t)$ with reflecting boundary conditions at $x=\pm a$ that started at $x(0)=0$. The main quantity of interest is the fraction of time
\begin{equation}\label{defS}
 S_T:=\frac{1}{T}\int_0^T\!\! dt\, \theta\big(x(t)\big)
\end{equation} 
the walker spent at positive values of $x$. Here $\theta$ denotes the Heaviside step function $\theta(x)=1$ for $x>1$ and $\theta(x)=0$ otherwise. 

With $x$ also $S_T$ is a random quantity. It is characterized by a probability density function $P_T(S)$ such that $P_T(S)dS$ gives the probability for $S\leq S_T\leq S+dS$.

The stochastic dynamics of the system are described by two coupled Langevin equations,
\begin{align}\label{LE1}
 \dot{x}&=\xi(t)\\\label{LE2}
 \dot{S}_T&=\frac{1}{T}\theta\big(x(t)\big),
\end{align}
where $\xi(t)$ denotes Gaussian white noise with expectation values 
\begin{equation}
 \lan \xi(t)\ran\equiv 0\spa{and}\lan\xi(t)\xi(t')\rangle=2D\delta(t-t').
\end{equation} 
The diffusion constant $D$ characterizes the noise strength. 

We measure $x$ in units of $a$ and rescale time such that $D=1/2$. The only remaining parameter in the problem is then the (dimensionless) time $T$. For $T\ll 1$ the walker has hardly a chance to feel the boundaries and $P_T(S)$ should be similar to the form shown in Fig.~\ref{fig:levy}. With $T$ increasing towards values of order one the boundary conditions become more and more relevant and the shape of $P_T(S)$ has to change accordingly. Finally, for large $T$ the walker has explored the whole interval $-1\leq x\leq 1$ evenly and we expect
\begin{equation}
 P_T(S)\to \delta\left(S-\frac{1}{2}\right).
\end{equation} 
The set of Langevin equations \eqref{LE1} and \eqref{LE2} is equivalent to the following Fokker-Planck equation for the joint probability density function $P(x,S,t)$ \cite{vanKampen}
\begin{equation}\label{FPE}
 \pa_t P(x,S,t)=-\frac{1}{T}\theta(x)\pa_S P(x,S,t)+\frac{1}{2} \pa_x^2 P(x,S,t).
\end{equation} 
This equation is complemented by zero-flux boundary conditions at $x=\pm 1$,
\begin{equation}\label{bc}
 \pa_x P(x,S,t)\Big|_{x=\pm 1}=0\qquad \forall t>0, \forall S\in[0,1],
\end{equation} 
and the initial condition
\begin{equation}\label{ic}
 P(x,S,0)=\delta(x)\delta(S).
\end{equation} 
From the symmetry of the problem it is clear that
\begin{equation}\label{symm}
 P(-x,1-S,t)=P(x,S,t).
\end{equation} 
Our central quantity of interest $P_T(S)$ is obtained from the solution $P(x,S,t)$ of the Fokker-Planck equation \eqref{FPE} by marginalization in $x$:
\begin{equation}\label{marg}
 P_T(S)=\int_{-1}^1\!\!dx\, P(x,S,T).
\end{equation}

%%%%%%%%%%%%%%%%%%%%%%%%%%%%%%%%%%%%%%%%%%%%%%%%%%%%%%%%%%%%%%%%%%%%%%%%%%%%%%%%%

\section{Solution of the Fokker-Planck equation}\label{sec:FPE}
Since $P(x,S,t)$ is defined on a finite interval of $S$ values it may be written as a Fourier series of the form 
\begin{equation}\label{FT}
 P(x,S,t)=\sum_p \psi_p(x,t)\, e^{2\pi ipS}.
\end{equation} 
The sum runs over all integer values of $p$ and the function $\psi_p(x,t)$ is given by
\begin{equation}\label{IFT}
 \psi_p(x,t)=\int_0^1\!\! dS\, P(x,S,t)\,e^{-2\pi ipS}.
\end{equation} 
Since $P(x,S,t)$ is real we have 
\begin{equation}\label{negp}
 \psi_{-p}(x,t)=\psi_p^*(x,t)
\end{equation} 
which together with \eqref{symm} results in 
\begin{equation}\label{symmpsi}
 \psi_p(x,t)=\psi_p^*(-x,t).
\end{equation} 

Multiplying \eqref{FPE} by $e^{-2\pi ipS}$ and integrating over $S$ we get
\begin{equation}\label{pdepsi}
 \pa_t \psi_p(x,t)=-\frac{2\pi i p}{T}\theta(x)\psi_p(x,t)+\frac{1}{2} \pa_x^2 \psi_p(x,t).
\end{equation} 
The boundary conditions translate to 
\begin{equation}\label{bcpsi}
 \pa_x \psi_p(x,t)\Big|_{x=\pm 1}=0\qquad \forall t>0, \forall p\in\mathbb{Z}
\end{equation} 
and the initial condition requires
\begin{equation}\label{ic1}
 \psi_p(x,0)=\delta(x)\qquad \forall p\in\mathbb{Z}.
\end{equation} 
Using the abbreviation 
\begin{equation}\label{defbp}
 \bp:=\frac{2\pi p}{T}
\end{equation} 
we solve \eqref{pdepsi} with the help of the separation ansatz
\begin{equation}\label{sepansatz}
 \psi_p(x,t)=e^{-\frac{1}{2}(E+i\bp)t}\,\ph(x).
\end{equation} 
Plugging \eqref{sepansatz} into \eqref{pdepsi} yields 
\begin{equation}\label{eveq}
 \ph''(x)+\big(E-i\bp\,\sign(x)\big)\ph(x)=0
\end{equation} 
subject to the boundary conditions
\begin{equation}\label{bcph}
 \ph'(x=\pm 1)=0.
\end{equation} 
Here $\sign(x)$ is the sign function
\begin{equation*}
 \sign(x)=\begin{cases}
           1 & x>0\\
          -1 & x\leq 0
          \end{cases},
\end{equation*}
and the prime denotes differentiation with respect to $x$. Alternatively, one may derive \eqref{eveq} from the path measure $P[x(\cdot)]$ using the Feynman-Kac formula, see Example 2 in \cite{Kac49}. From \eqref{symmpsi} we have 
\begin{equation}\label{symmph}
 \ph(x)=\ph^*(-x).
\end{equation} 

We expect for each $p$ a denumerable set of discrete eigenvalues $E^{(p)}_n$ and corresponding eigenfunctions $\ph^{(p)}_n(x)$ solving \eqref{eveq}. Standard arguments show that the spectrum is non-degenerate and that eigenfunctions corresponding to different eigenvalues are orthogonal:
\begin{equation}\label{defa}
 \int_{-1}^1\!\!dx\,\ph^{(p)}_n(x)\,\ph^{(p)}_m(x)=:a^{(p)}_n\,\delta_{nm}.
\end{equation}
The general solution to eq.~\eqref{pdepsi} is hence of the form
\begin{equation}\label{solpsi}
 \psi_p(x,t)=\sum_{n=1}^\infty c^{(p)}_n\,\ph^{(p)}_n(x) \,e^{-\frac{1}{2}(E^{(p)}_n+i\bp)t}.
\end{equation} 
It has to be kept in mind that due to the imaginary term $i\bp\,\sign(x)$ in eq.~\eqref{eveq} $E^{(p)}_n$ and $\ph^{(p)}_n(x)$ and consequently also the normalization and expansion coefficients $a^{(p)}_n$ and $c^{(p)}_n$, respectively, will in general be complex.

The $c^{(p)}_n$ may be determined from the initial condition \eqref{ic1}:
\begin{equation}
 \delta(x)=\psi_p(x,0)=\sum_{n=1}^\infty c^{(p)}_n\,\ph^{(p)}_n(x). 
\end{equation}
Multiplying with $\ph^{(p)}_m(x)$, integrating over $x$ and using the orthogonality \eqref{defa} yields
\begin{equation}\label{defd}
d_m^{(p)}:= \ph^{(p)}_m(0)=c_m^{(p)}\,a_m^{(p)}
\end{equation}

Moreover, in view of \eqref{marg} we do not need the complete function  $\psi_p(x,t)$ to finally determine $P_T(S)$. It is sufficient to know
\begin{equation}\label{defchi}
 \chi_p(T):=\int_{-1}^1\!\! dx \, \psi_p(x,T)
\end{equation}
from which we get using \eqref{FT} 
\begin{equation}\label{FTchi}
 P_T(S)=\sum_p \chi_p(T)\, e^{2\pi ipS}.
\end{equation} 
Defining 
\begin{equation}\label{defb}
 b_m^{(p)}:=\int_{-1}^1\!\! dx\, \ph^{(p)}_m(x),
\end{equation}
combining \eqref{solpsi} with \eqref{defd} and \eqref{defb}, and observing \eqref{defbp} we end up with
\begin{equation}\label{solchi}
 \chi_p(T)= (-1)^p\,
    \sum_{n=1}^\infty \frac{b^{(p)}_n d^{(p)}_n}{a^{(p)}_n}\,e^{-\frac{1}{2} E^{(p)}_nT}.
\end{equation}
Whenever no confusion may arise we will suppress the superscript $p$ at $E,\ph(x),a,b$ and $d$ in the following to lighten the notation.

%%%%%%%%%%%%%%%%%%%%%%%%%%%%%%%%%%%%%%%%%%%%%%%%%%%%%%%%%%%%%%%%%%%%%%%%%%%%%%%%%%%%%

\section{The eigenvalue problem}\label{sec:SL}
To complete the determination of $P_T(S)$ via \eqref{solchi} and \eqref{FTchi} we need to solve the eigenvalue problem \eqref{eveq}
\begin{equation}\label{evp}
 \ph_n''(x)+\big(E_n-i\bp\,\sign(x)\big)\ph_n(x)=0
\end{equation} 
for all integer values of $p$. The real part of $E_n$ must always be nonnegative. To see this we multiply \eqref{evp} with $\ph_n^*(x)$, integrate over $x$ and take the real part to find 
\begin{equation}
 \Re(E_n)\int_{-1}^1\!\!dx\,|\ph_n(x)|^2=\int_{-1}^1\!\!dx\,|\ph'_n(x)|^2.
\end{equation}
The integral on the l.h.s. of this equation must be positive for $\ph_n(x)$ to be an eigenfunction, the one on the r.h.s. is nonnegative. This proves the assertion. Similar arguments show that $E=0$ is possible only for $\bp=0$. 

\begin{figure}
\centering
\includegraphics[width=.6\linewidth]{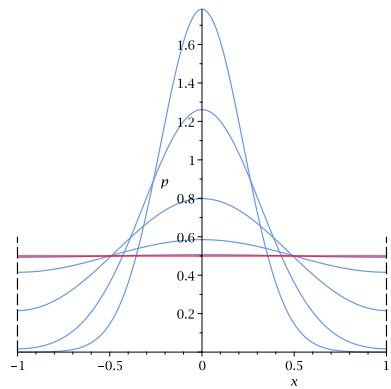}
\caption{Blue lines show the probability density function $p(x,t)$ for the position $x$ of the walker for $t=0.05, 0.1, 0.25, 0.5$ and $t=1$ (middle top to bottom). The red line is the stationary distribution $p_\mathrm{st}(x)\equiv 1/2$ reached for $t\to\infty$.}
\label{fig:relax}
\end{figure}

In fact, for $p=0$ the whole eigenvalue problem is equivalent to a standard exercise in quantum mechanics \cite{LLIIIpar22} and may be solved analytically. The result reads 
\begin{equation}\label{resp=0}
 \psi_0(x,t)=\frac{1}{2}+\sum_{n=1}^\infty \cos(n\pi x)\,e^{-\frac{1}{2}n^2\pi^2 t}.
\end{equation}
Note that from \eqref{IFT} it follows that 
\begin{equation}
 \psi_0(x,t)=\int_0^1\!\!dS \,P(x,S,t)=:p(x,t).
\end{equation}
Hence, \eqref{resp=0} describes the time-evolution of the probability density function $p(x,t)$ for the position of the walker. Fig.\ref{fig:relax} shows a few snapshots. Moreover, Eq.~\eqref{defchi} implies 
\begin{equation}
 \chi_0(t)\equiv 1
\end{equation} 
which via \eqref{FTchi} ensures the normalization of $P_T(S)$ for all $T$.

\begin{figure*}
\includegraphics[width=.25\linewidth]{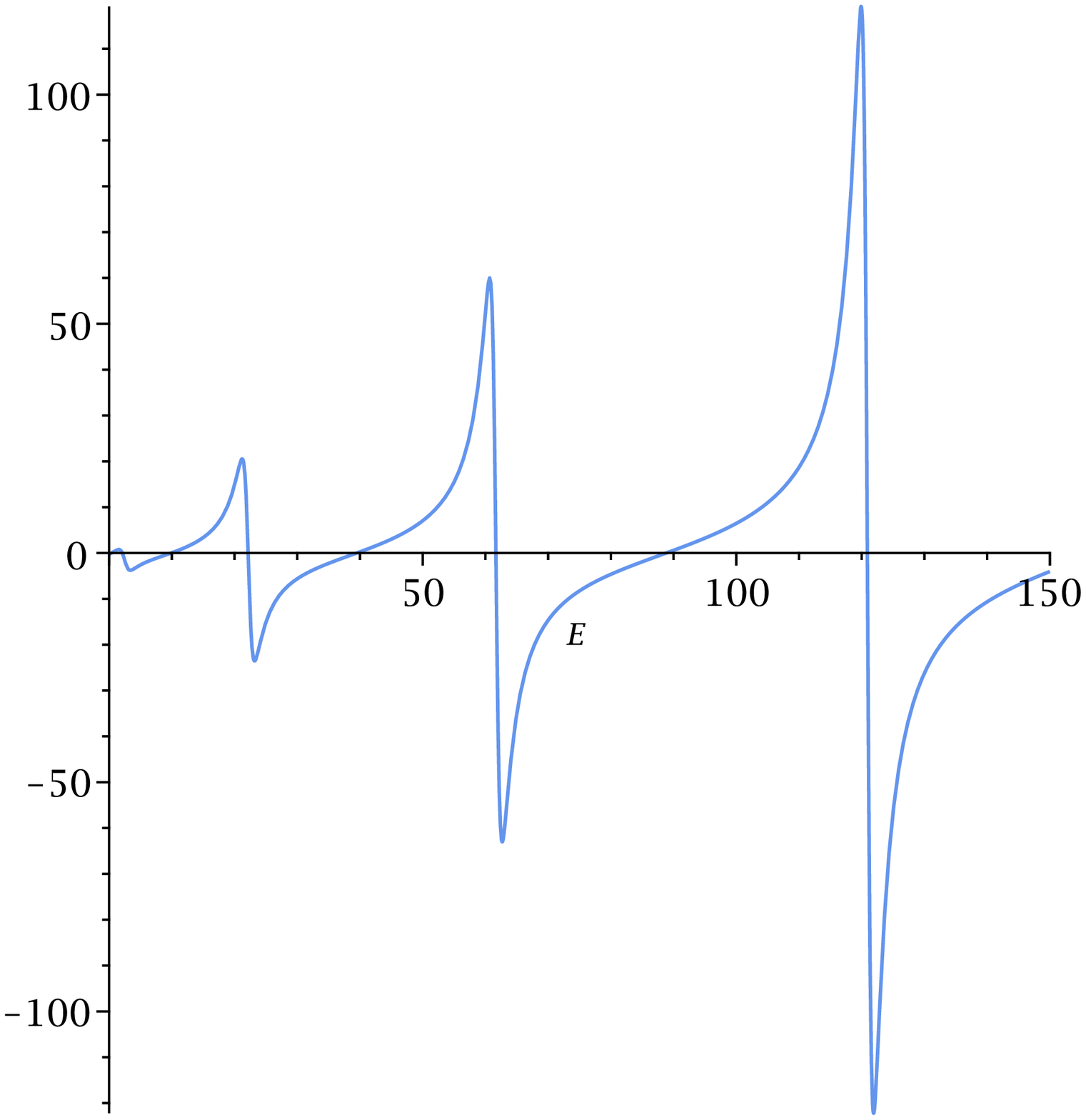}\hspace{4ex}
\includegraphics[width=.25\linewidth]{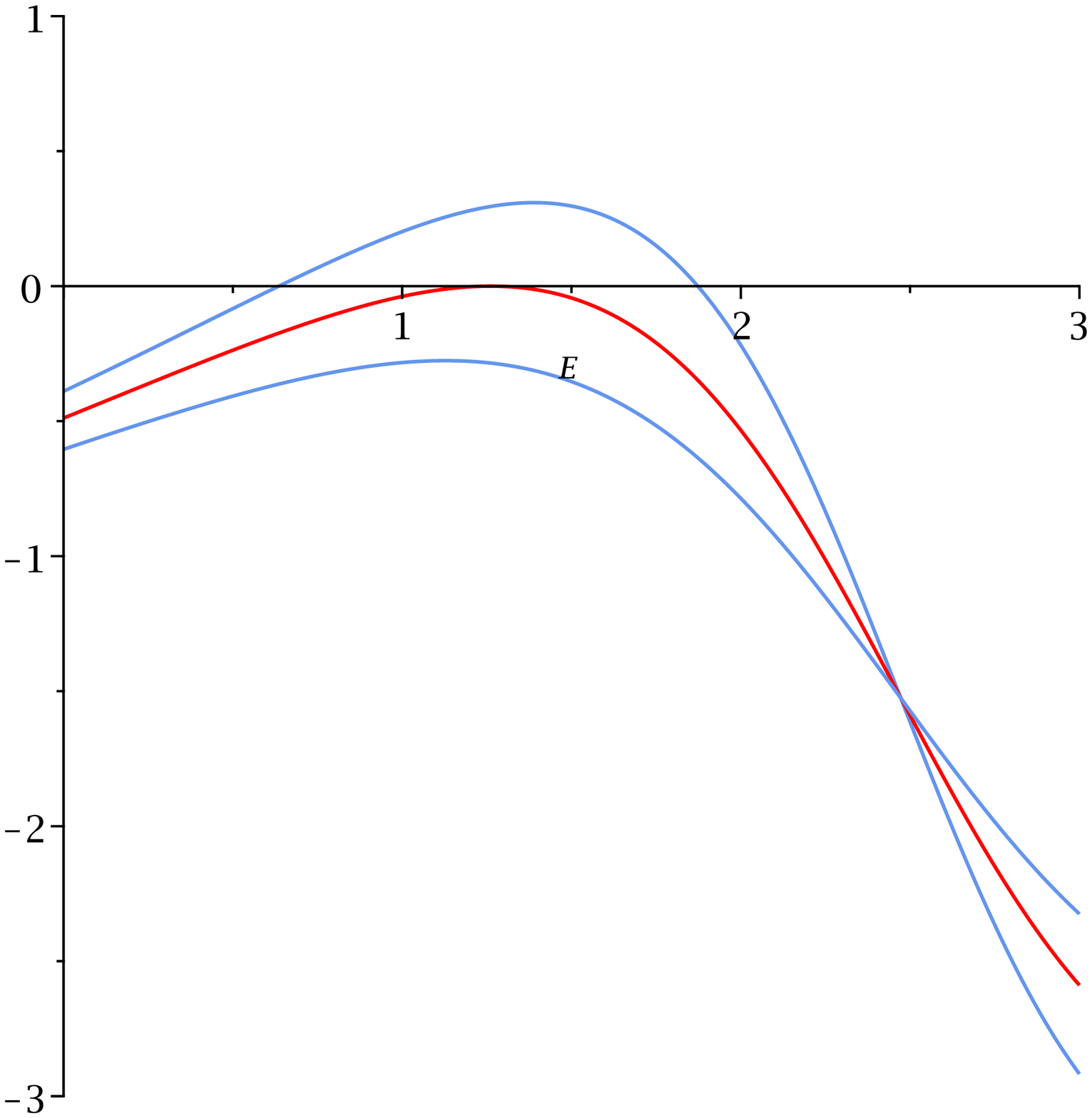}\hspace{4ex}
\includegraphics[width=.25\linewidth]{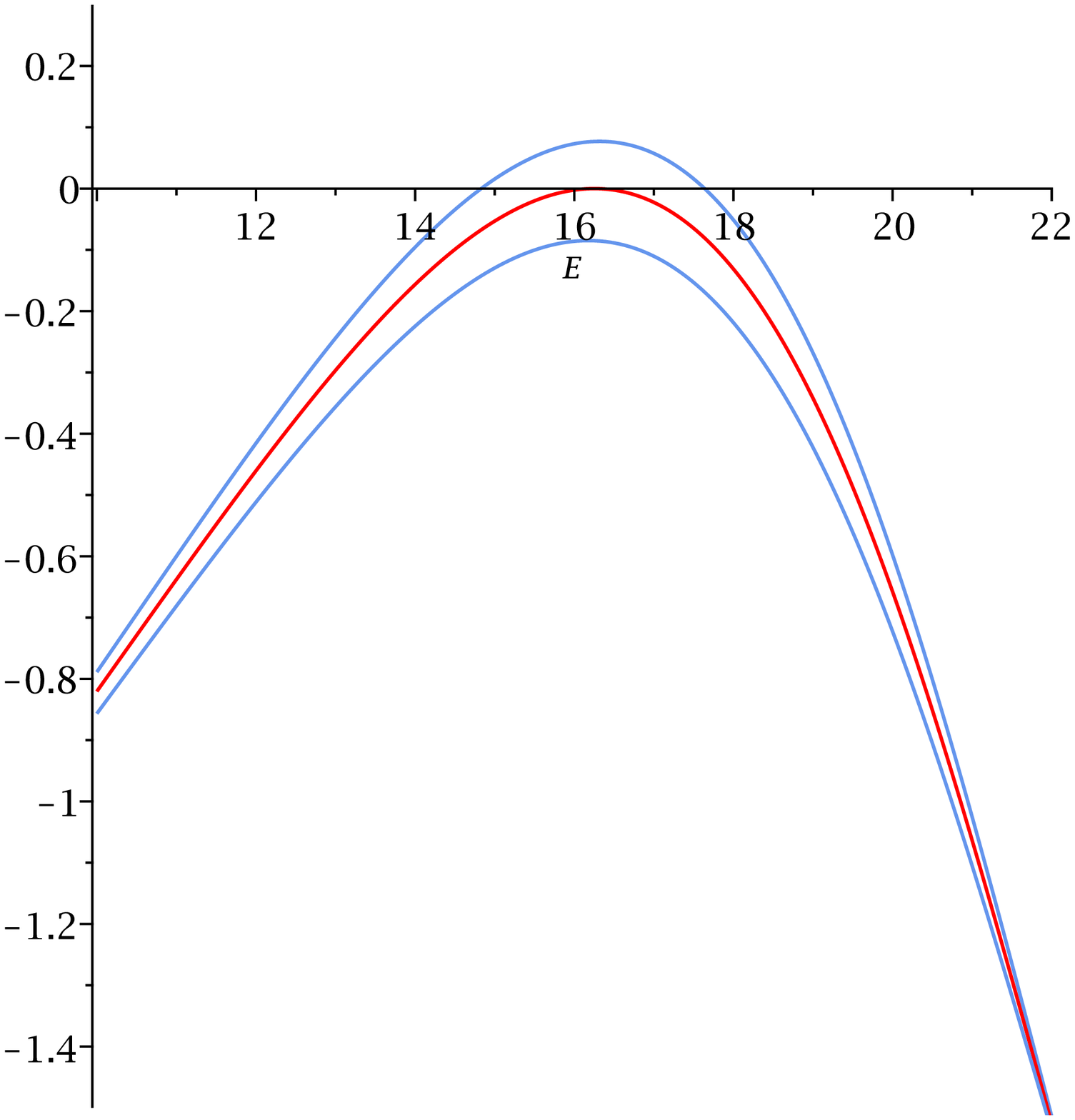}
\caption{Left: Left hand side of equation \eqref{eqE} for real $E$ and $\bp=1$. Middle: Blowup for small values of $E$ and $\bp=1.2, 1.3855, 1.6$ (left top to bottom). Right: Blowup around the second pair of roots for $\bp=8.2, 8.3862, 8.6$ (left top to bottom).\label{fig:eqE}}
\end{figure*}

For $p\neq 0$ \eqref{evp} is a linear ordinary differential equation with piecewise constant coefficients. It is solved for $x<0$ and $x>0$ separately and the solutions are then matched at $x=0$ such that $\ph_n$ and $\frac{d}{dx}\ln\ph_n$ are continuous there. With the abbreviations
\begin{equation}\label{defkq}
 k_n:=\sqrt{E_n+i\bp}\spa{and} q_n:=\sqrt{E_n-i\bp}
\end{equation} 
the result is
\begin{equation}\label{resph}
 \ph_n(x)=\begin{cases}
           \ds\sqrt{\frac{\cos q_n}{\cos k_n}}\, \cos k_n(x+1) & -1\leq x\leq 0\\
           \ds\sqrt{\frac{\cos k_n}{\cos q_n}}\, \cos q_n(x-1) &  0\leq x\leq 1\\ 
          \end{cases}
\end{equation} 
where $E_n$ has to fulfill 
\begin{equation}\label{eqE}
 \sqrt{E_n+i\bp}\,\tan \sqrt{E_n+i\bp}+\sqrt{E_n-i\bp}\,\tan \sqrt{E_n-i\bp}=0.
\end{equation} 
The prefactors of the cosine functions in \eqref{resph} have been chosen such that \eqref{symmph} holds. From \eqref{resph} the coefficients $a_n, b_n$ and $d_n$ can be determined from their respective definitions \eqref{defa}, \eqref{defb}, \eqref{defd}. We find 
\begin{align}\label{resbd}
b_n\,d_n&=\frac{\cos q_n \sin k_n}{k_n}+\frac{\cos k_n \sin q_n}{q_n}\\\label{resa}
a_n&=\frac{1}{2}\left(\frac{\cos q_n}{\cos k_n}+\frac{\cos k_n}{\cos q_n}+b_n\,d_n\right).
\end{align}

%%%%%%%%%%%%%%%%%%%%%%%%%%%%%%%%%%%%%%%%%%%%%%%%%%%%%%%%%%%%%%%%%%%%%%%%%%%%%%%%%%%%%%%%%%%%%%%

\section{Numerical determination of the eigenvalues}\label{sec:num}

Eq.~\eqref{eqE} can only be solved numerically. Nevertheless a few prior consideration are in order. 
Taking the complex conjugate of \eqref{evp} and using the fact that $\sign(x)$ is an odd function of $x$ we see that with $E_n,\ph_n(x)$ also $E^*_n,\ph_n^*(-x)$ is an admissible solution. Complex eigenvalues hence come in pairs of conjugates entailing the same for the associated constants $a_n, b_n$ and $c_n$. 

Moreover, for real $E_n$ the imaginary part of eq.~\eqref{eqE} is identically zero \cite{rem1}. Fig.~\ref{fig:eqE} shows plots of the l.~h.~s. of~\eqref{eqE} for real $E$. From the left plot we infer that the gap between successive real roots $E_n$ is much larger than 1. Keeping in mind that we are interested in values of $T$ of order 1 only the contributions from the first two roots will play a noticeable role in the superposition \eqref{solchi}. This is corroborated by comparison with our numerical simulations, see section~\ref{sec:res} below. The higher eigenvalues are only important for the very short time dynamics in which $P_T(S)$ has to transform from the initial condition $\delta(S)$ to the shape shown in Fig.~\ref{fig:levy}.

\begin{figure}
\centering
\includegraphics[width=.6\linewidth]{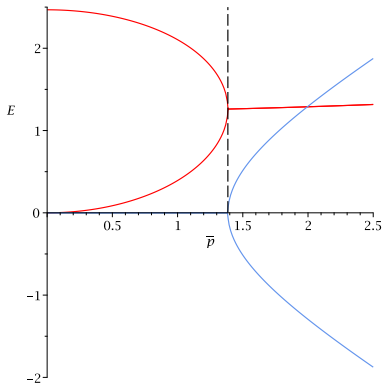}
\caption{The two solutions of the eigenvalue equation \eqref{eqE} with the smallest real part as function of $\bp$. The real part is shown in red, the imaginary one in blue. At $\bp=\bp_{\mathrm{c1}}\simeq 1.38$ the two real solutions merge and give birth to a pair of complex conjugated solutions.} 
\label{fig:eqE4}
\end{figure}

The other two plots of Fig.~\ref{fig:eqE} show that with increasing $\bp$ the real roots disappear successively to give way to pairs of complex conjugate solutions for $E_n$. At the critical values $\bp_\mathrm{ci}$ of $\bp$ where these transitions occur the l.~h.~s. of \eqref{eqE} and its derivative with respect to $E_n$ both vanish. For the first two bifurcation values we find  $\bp_{\mathrm{c1}}=1.385577425$ and $\bp_\mathrm{c2}=8.386237648$, respectively. The first bifurcation of this type is shown in detail in Fig.~\ref{fig:eqE4}.

\begin{figure*}
\centering
\includegraphics[width=.25\linewidth]{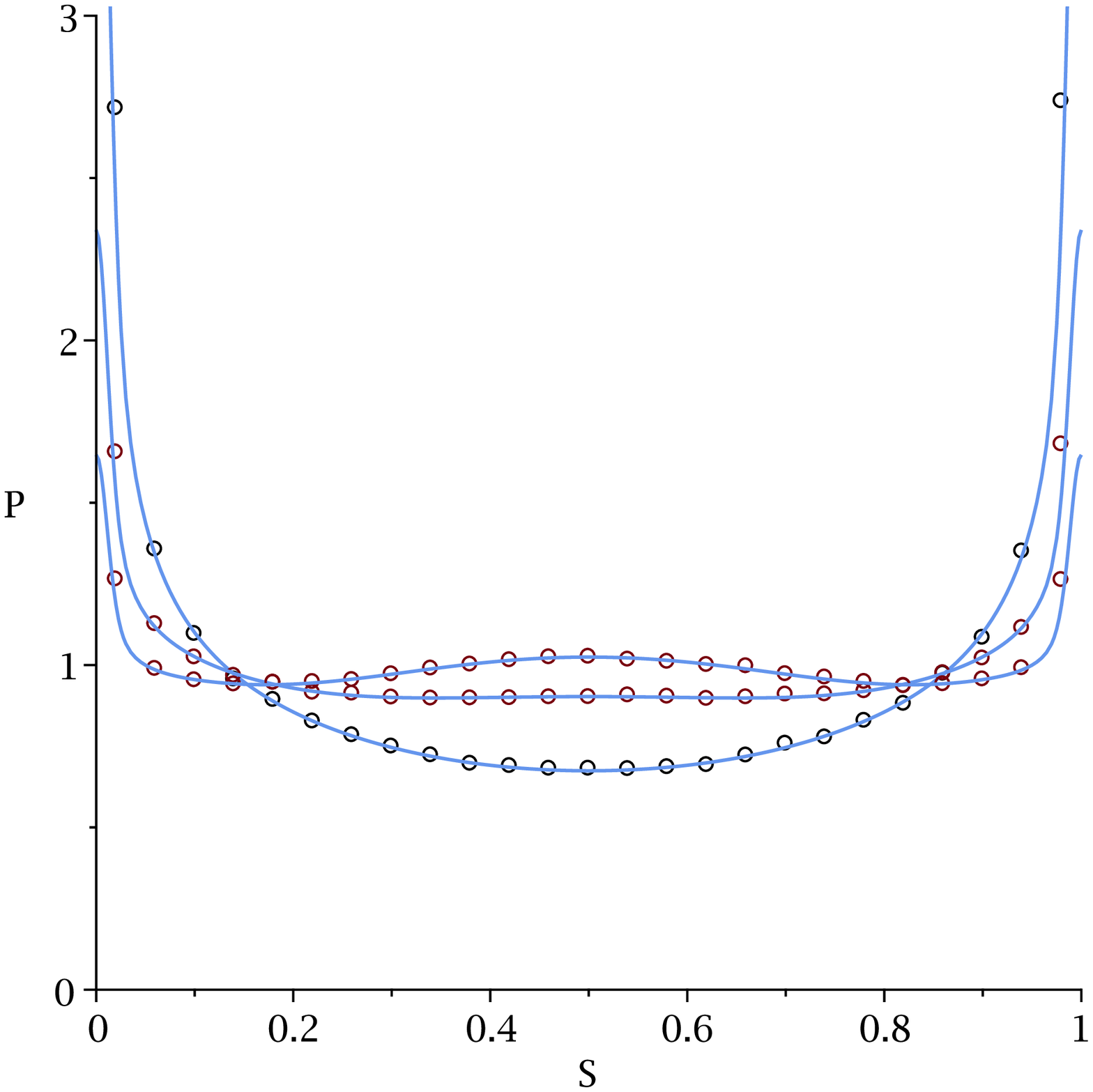}\hspace{4ex}
\includegraphics[width=.25\linewidth]{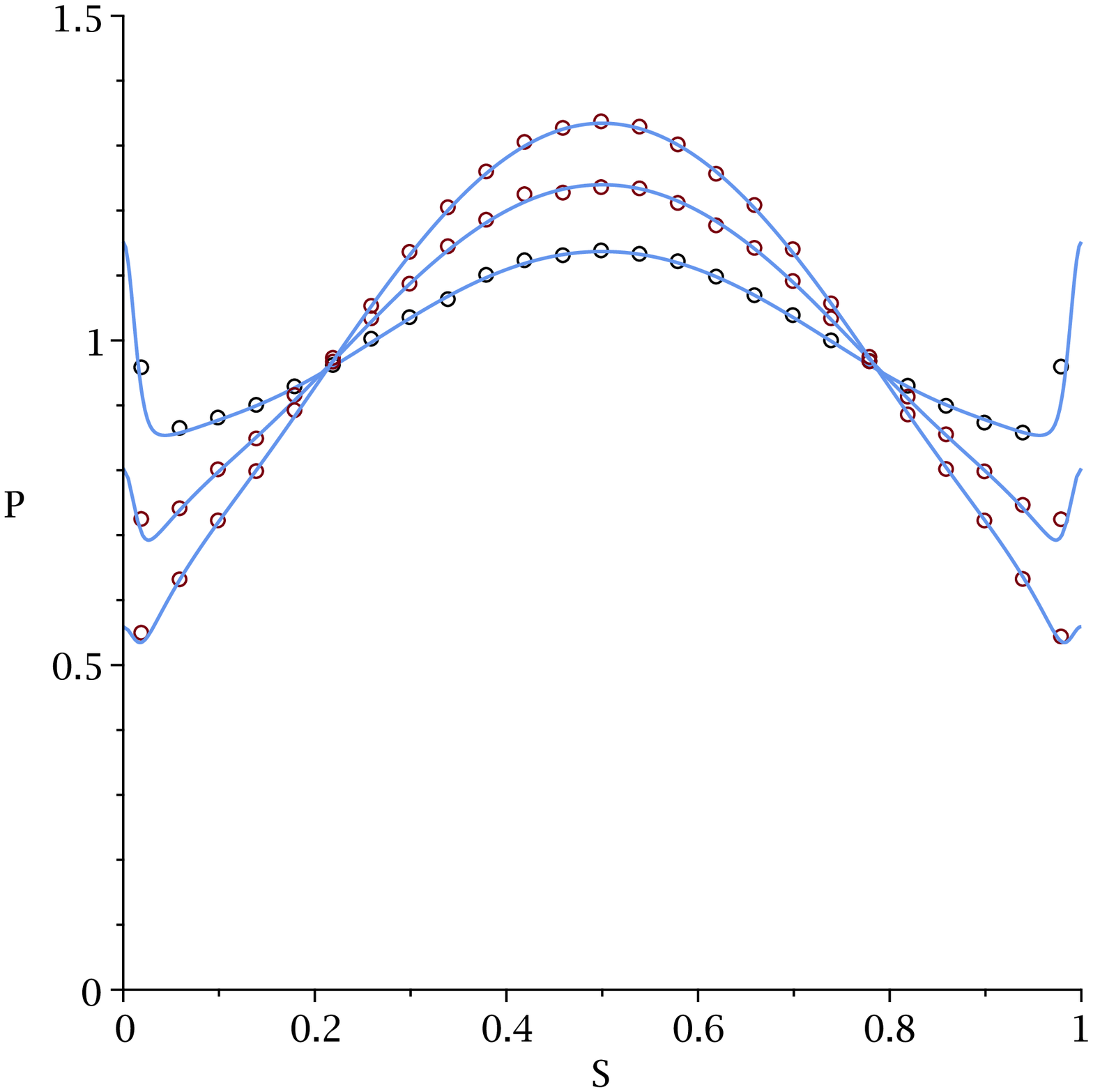}\hspace{4ex}
\includegraphics[width=.25\linewidth]{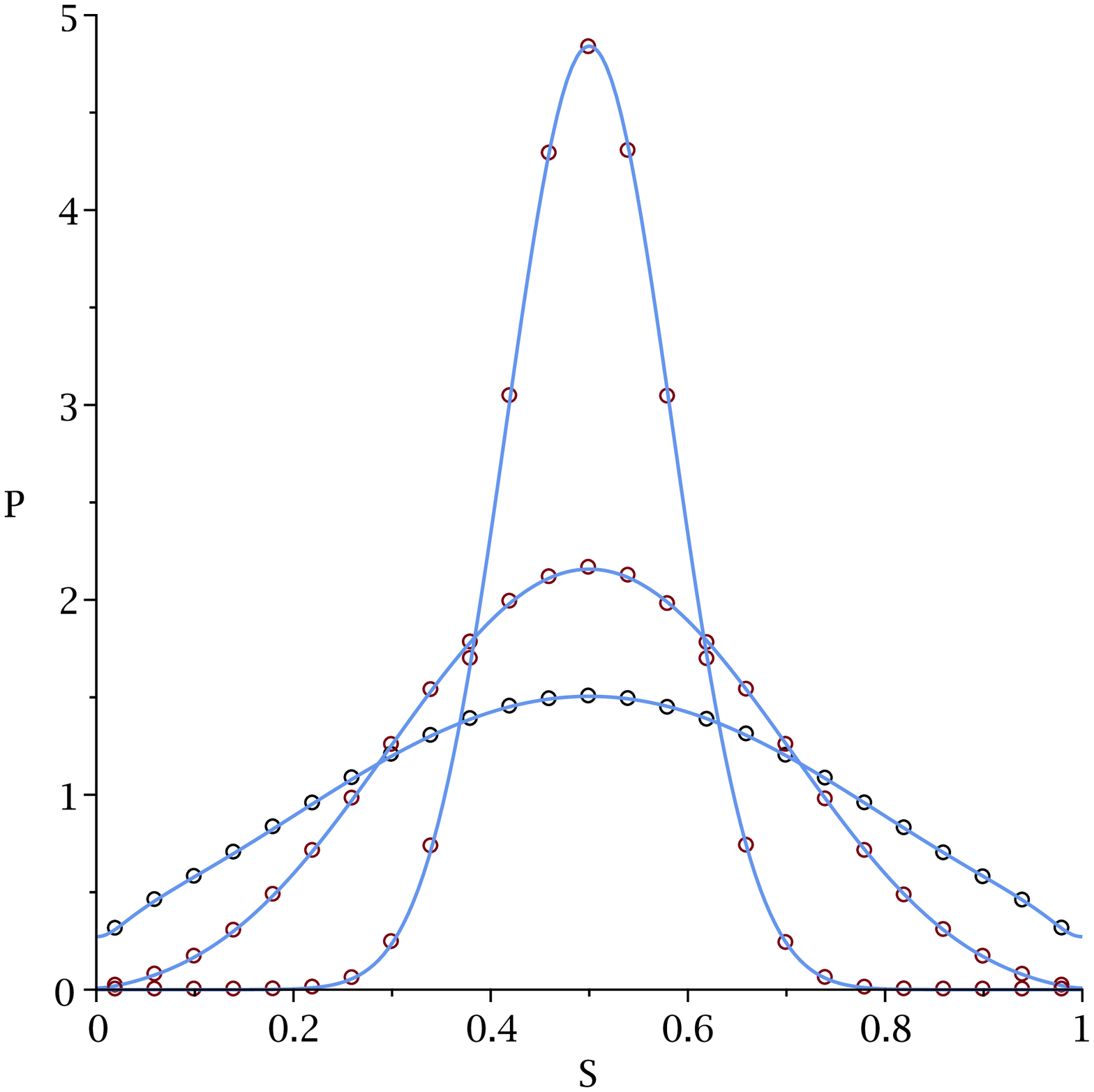}
\caption{Blue lines are plots of $P_T(S)$ obtained as described in the main text. Symbols are results of numerical simulations with time step $\Delta t=5\cdot 10^{-4}$ averaged over $2.5\cdot 10^6$ realizations. Statistical errors are smaller than the symbol size. Left: $T=1,2,2.5$ (middle bottom to top). Middle: $T=3,3.5,4$ (middle bottom to top); note the reduced scale for $P_T$. Right: $T=5,10, 50$ (middle bottom to top).\label{fig:res}}.
\end{figure*}

These considerations make clear how to get rather accurate approximate results for $P_T(S)$. First, depending on the desired resolution for $S$ the maximal number $p_\mathrm{max}$ of Fourier modes is chosen. Then for given value of $T$ one checks for each $p$ whether $\bp$ is smaller or larger than $\bp_\mathrm{c1}$. For $\bp<\bp_\mathrm{c1}$ one determines the two lowest real solutions $E_1$ and $E_2$ numerically from \eqref{eqE} and calculates the corresponding values of $a, b$, and $d$. If $\bp>\bp_\mathrm{c1}$ it is sufficient to determine one eigenvalue $E_1$ with its coefficients $a, b$, and $d$; $E_2$ and its corresponding coefficients then follow by complex conjugation. Because of \eqref{negp} it is sufficient to consider positive values of $p$ only. Plots obtained in this way are shown in section \ref{sec:res} below together with results from numerical simulations. 

\section{Asymptotics for large $T$}\label{sec:asy}
The asymptotics of $\chi_p(T)$ as given by \eqref{solchi} for large $T$ is not completely obvious since $T$ enters the eigenvalue equation \eqref{eqE} via $\bp$, cf. Eq.~\eqref{defbp}, and therefore the $E_n$ depend on $T$. For $T\to\infty$ we have $\bp\to 0$ giving rise to $E_1\to 0$. An expansion of \eqref{eqE} in $\bp$ and $E_1$ yields the asymptotic behaviour
\begin{equation}\label{asyE}
 E_1\sim\frac{\bp^2}{3}.
\end{equation}
For large $T$ we therefore have $E_1=O(1/T^2)$ and consequently $E_1\,T\to 0$. At the same time we find from \eqref{resbd} and \eqref{resa} for $\bq\to 0$ by using \eqref{defkq} and \eqref{asyE} 
\begin{equation}
 \frac{b_1\,d_1}{a_1}\to 1.
\end{equation} 
Hence \eqref{solchi} results in $\zeta_p(T)\to (-1)^p$ for $T\to\infty$ which implies 
\begin{equation}\label{asyP}
 P_T(S)\to\delta\left(S-\frac{1}{2}\right)\spa{for} T\to \infty
\end{equation} 
as expected. 

%%%%%%%%%%%%%%%%%%%%%%%%%%%%%%%%%%%%%%%%%%%%%%%%%%%%%%%%%%%%%%%%%%%%%%%%%%%%%%%%%%%%%%%

\section{Results}\label{sec:res}
Results for $P_T(S)$ obtained along the lines of sections~\ref{sec:FPE}-\ref{sec:num} for different values of $T$ are shown in Fig.~\ref{fig:res}. We have chosen $p_\mathrm{max}=128$ as number of Fourier modes which is sufficient to resolve the important details of $P_T(S)$. To suppress spurious oscillations in particular in the almost constant parts of $P_T(S)$ for small $T$ we have additionally smoothed the results with a Gaussian filter of width $\sigma=0.01$.

The left plot of Fig.~\ref{fig:res} shows that for $T\lesssim 2$ there are only small modifications in $P_T(S)$ as compared to the case without boundary conditions shown in Fig.~\ref{fig:levy}. Only few realizations are able to reach the boundaries at $x=\pm 1$ and to get back and cross the starting point to contribute to the slight increase of $P_T(S)$ near $S=1/2$. Note that, in marked contrast, for these values of $T$ the distribution of the walker itself is already near to the stationary state, cf. Fig.~\ref{fig:relax}. The equilibration of $p(x,t)$ therefore occurs mostly separately in the regions of positive and negative $x$ without many crossings of the starting point. 

For $2.5\leq T\leq 4$ the main reshaping of $P_T(S)$ from a bimodal distribution with maxima at $S=0,1$ to a unimodal one with maximum at $S=1/2$ takes place. This is demonstrated by the middle part of Fig.~\ref{fig:res}. For the values of $T$ shown the walker typically not only reaches one of the boundaries but also has enough time to get back to the starting point. As a result more and more of the realizations cross the starting point and contribute to the growing central maximum of $P_T(S)$. At the same time it becomes increasingly unlikely for the walker to stay in only the left or the right half of the allowed interval resulting in a steady decrease of $P_T(S)$ near the boundary values $S=0$ and $S=1$.

Finally, for $T\geq 5$ most realizations have visited both boundaries and crossed the origin several times. Accordingly, $P_T(S)$ approaches a Gaussian centered at $S=1/2$ that becomes sharper with increasing $T$ to eventually approach the limit given by \eqref{asyP}. The right part of Fig.~\ref{fig:res} shows some intermediate stages. 

All results obtained are in excellent agreement with numerical simulations shown by the symbols in Fig.~\ref{fig:res}. This a posteriori validates the restriction of the expansion \eqref{solchi} to the two leading  eigenvalues.

\section{Conclusion} \label{sec:conc}
In conclusion we have shown that the somewhat surprising shape of the distribution $P_T(S)$ for the fraction $S$ of total time $T$ an unrestricted random walker spends to the right of the starting point as shown in Fig.~\ref{fig:levy} is indeed due to the possibility of very large excursions away from this point. When restricting the walker to a finite interval by reflecting boundary conditions these excursions are precluded and $P_T(S)$ assumes a unimodal form with maximum at $S=1/2$ for sufficiently large $T$. Unlike the case without boundaries no complete analytical solution seems possible. However, rather accurate approximate results may be obtained on the basis of just the two leading eigenvalues of the corresponding Fokker-Planck operator. The results are in excellent agreement with numerical simulations, and the emerging picture is consistent with physical intuition. 

\begin{acknowledgments}
 We are grateful to Eli Barkai and Hugo Touchette for pointing out pertinent references.
\end{acknowledgments}

\bibliography{references}

\begin{thebibliography}{99}
\bibitem{rem1} This is one of the virtues of separating $i\bp$ from $E$ in the ansatz \eqref{sepansatz}.
\end{thebibliography}

\end{document}